\documentclass[aps,pra,twocolumn,showpacs,amsmath,
amssymb,amsfonts,superscriptaddress]{revtex4}
\usepackage{graphics}
\usepackage{epsfig}

\begin{document}

\title{Privacy of a lossy bosonic memory channel}

\author{Giovanna Ruggeri}
\email{ruggeri@le.infn.it}
\affiliation{Dipartimento di Fisica, 
Universit\`a di Lecce, I-73100 Lecce, Italy}

\author{Stefano Mancini}
\email{stefano.mancini@unicam.it}
\affiliation{Dipartimento di Fisica, Universit\`{a} di Camerino, 
I-62032 Camerino, Italy}

\begin{abstract}
We study the security of the information transmission between two honest parties realized through a lossy bosonic \emph{memory} channel when losses are captured by a dishonest party. We then show that entangled inputs can enhance the private information of such a channel, which however does never overcome that of unentangled inputs in absence of memory.
\end{abstract}


\maketitle

\section{Introduction}

Quantum communications concern the study of quantum channels that also use continuous alphabets \cite{brau}.
These can be modeled by bosonic field modes whose phase space quadratures enable for continuous variable encoding/decoding. 
Lossy bosonic channel, which consists of a collection of bosonic modes that lose energy en route from the transmitter to the receiver, belongs to the class of Gaussian channels which provide a fertile testing ground for the general theory of quantum channels' capacities \cite{Hol99} and  are easy to implement experimentally with high accuracy level (beam splitters or squeezers are examples of Gaussian operations) \cite{brau}.
Hence, the increasing attention devoted to channels with memory effects, where
the noise may be strongly correlated between uses of the channel, 
has been extended to bosonic channels \cite{GM05}.
The main motivation that has led to investigate memory effects in such channels
has been the possibility to enhance their classical capacity by means of entangled inputs 
\cite{Rug05, Cerf04}. Thus, the quest for effectiveness of entanglement inputs in other kind of capacities.
Here we study the  
\emph{private classical information capacity} \cite{Dev}
 of a lossy bosonic memory channel.
We use the model introduced in Ref.\cite{Rug05} where the memory effects are realized by considering quantum correlations among environments acting on different channel uses \cite{GM05}. We then analyze the security (privacy) of the channel when there is a third dishonest party, who captures the information lost during the process of transmission between the sender and the receiver, i.e. the eavesdropper accesses the environment's final state.
We shall show that entangled inputs can enhance the private classical information capacity \cite{Dev}, but not above that of unentangled inputs in absence of memory. This is in contrast with what happen for simple classical information capacity \cite{Rug05,Cerf04}.


\section{The model}

Let us consider a $2$-shot  bosonic channel with correlated noise, acting on two independent modes of the electromagnetic field associated with the annihilation operators  ${\hat a}_1,{\hat a}_2$. Here, each mode ${\hat a}_k$ (with  $k=1,2$) interacts with an environment mode ${\hat b}_k$ through a beam splitter of transmittivity $\eta\in[0,1]$, modeling losses.

For single mode Gaussian channels it is conjectured
that Gaussian inputs suffice to achieve the classical capacities. Here we do not question this generally admitted conjecture coming form Ref.\cite{Hol99}, and we assume it to be also valid for two-mode (memory) channels, likewise Refs.\cite{Cerf04}. 
Hence, we consider a general Gaussian input as a mixture of entangled (two-mode squeezed) coherent states, given by
 \begin{equation}
 |\psi \left( \mu_1\,,\mu_2 \right)\rangle={\hat S}_{a}(r)\left [{\hat D}_{a_2} (\mu_2)|0 \rangle_{a_2} {\hat D}_{a_1}(\mu_1)|0\rangle_{a_1}\right],
 \label{psi}
 \end{equation}
where ${\hat D}_{a_k} (\mu_k)=\exp \left(\mu_k {\hat a}_{k}^{\dagger}-\mu_k^* {\hat a}_{k} \right)$ is the single $k$-mode displacement operator, corresponding to the complex number $\mu_k$
 and ${\hat S}_a(r)=\exp \left[\frac{1}{2}r \left( {\hat a}_1^{\dagger}{\hat a}_2^{\dagger}-{\hat a}_1 {\hat a}_2\right) \right]$ denotes the two-mode squeeze operator \cite{brau},
 with $r$ the entanglement parameter between the two inputs ($r=0$ refers to input  product states).
 Let us write the input (\ref{psi}) as a density operator
\begin{eqnarray}
 {\rho}_{in}\left(\mu_1\,,\mu_2 \right)=|\psi\left(\mu_1\,,\mu_2 \right)\rangle \langle\psi\left(\mu_1\,,\mu_2 \right)|.
\label{rhoin}
\end{eqnarray}
Then, we assume such states weighted with the Gaussian probability distribution
\begin{equation}
P\left(\mu_1\,,\mu_2 \right)=\frac{1}{\pi^2 N^2} \exp  \left( -\frac{|\mu_1|^2+|\mu_2|^2}{N}\right)\,,
\label{Pimu}
\end{equation}
where $N$ is the average photon number per channel use.
In practice, due to entanglement, the effective average photon number per channel use will be $N_{eff}=N+\sinh^2 r$. 

According to Ref.\cite{GM05}, we introduce the correlations between the environment actions on the two channel uses by a two-mode squeezed vacuum state,
 \begin{equation}
 {\rho}_{env}={\hat S}_b(s)\left [|0 \rangle_{b_2}|0\rangle_{b_1}{}_{b_1}\langle 0|{}_{b_2}\langle 0|\right] {\hat S}^{\dag}_b(s)
 \label{rhoenv}
 \end{equation}
where  ${\hat S}_b(s)=\exp \left[\frac{1}{2}s \left( {\hat b}_1^{\dagger}{\hat b}_2^{\dagger}-{\hat b}_1 {\hat b}_2\right) \right]$
 with $s$ the memory parameter ($s=0$ corresponds to memoryless case).
In our model, the interaction between input and environment is characterized by the beam splitter transformation \cite{brau}
\begin{eqnarray}
  \label{unouno}
  {\hat a}_k&\mapsto&\sqrt{\eta}\;  {\hat a}_k-\sqrt{1-\eta}\;  {\hat b}_k,\\
  {\hat b}_k&\mapsto&\sqrt{1-\eta}\;  {\hat a}_k+\sqrt{\eta}\;  {\hat b}_k. 
  \label{duedue}
  \end{eqnarray}  
As consequence the input and environment states are mapped onto a global (possibly entangled) state  ${\rho}$
\begin{equation}
{\rho}_{in}\otimes{\rho}_{env}\mapsto {\rho}.
\label{rhomap}
\end{equation}  
Then, on one hand, the receiver will get the output state $\rho_{out}={\rm Tr}_{b_1b_2}(\rho)$.
On the other hand, the losses might be captured by a dishonest party, the eavesdropper, accessing
the environment's final state ${\rho}_{eve}={\rm Tr}_{a_1a_2}(\rho)$.

Since we deal with Gaussian states, it turns useful to work with Wigner distribution functions \cite{brau}.
Let $q_k\,,p_k$ be the quadrature variables of $\hat{a}_k$, i.e. the classical variable corresponding to 
$\hat{q}_k=(\hat{a}_k +\hat{a}_k^{\dag})/\sqrt{2}$, $\hat{p}_k=-i(\hat{a}_k -\hat{a}_k^{\dag})/\sqrt{2}$.
Moreover, let $\mu_k^R\,,\mu_k^I$ be the real and imaginary part of $\mu_k$.
We introduce the row vectors  in $\mathbb{R}^{4}$,
 \begin{eqnarray} 
\mbox{\boldmath$u$}&=&\left(
 q_1,q_2,p_1,p_2\right)\,,\\
 \mbox{\boldmath$\mu$}&=&\left(
 \mu_1^R,\mu_2^R,\mu_1^I,\mu_2^I\right)\,,
\end{eqnarray}  
and the real $4 \times 4$ matrix
\begin{equation}
\label{Ar}
 {\cal A}_r=
 \left(
 \begin{array}{cccc}
 \cosh {2r} &  -\sinh {2r} & 0 & 0 \\
 -\sinh {2r}  & \cosh {2r} & 0 & 0 \\
  0 & 0 & \cosh {2r} &  \sinh {2r} \\
  0 & 0 &  \sinh {2r}  & \cosh {2r}
\end{array}
 \right)\,.
\end{equation}
The  Wigner function corresponding to ${\rho}_{in}$ in Eq.(\ref{rhoin}) reads 
\begin{equation} 
 W_{in}(\mbox{\boldmath$u$};\mbox{\boldmath$\mu$})=\frac{1}{\pi^2}
 \exp\left[-\mbox{\boldmath$u$}{\cal A}_r \mbox{\boldmath$u$}^T-\mbox{\boldmath$\mu$}{\cal A}_r \mbox{\boldmath$\mu$}^T+2\mbox{\boldmath$\mu$}{\cal A}_r \mbox{\boldmath$u$}^T\right].
\label{Win}
\end{equation}
Moreover, let us denote by
\begin{eqnarray} 
 \mbox{\boldmath$v$}&=&\left(x_1,x_2, y_1,y_2\right)\,
 \end{eqnarray}
 the real  4-component vector of  the quadrature variables $x_k$, $y_k$ associated with the environment operators  ${\hat b}_k$, i.e. the classical variable corresponding to 
$\hat{x}_k=(\hat{b}_k +\hat{b}_k^{\dag})/\sqrt{2}$, $\hat{y}_k=-i(\hat{b}_k -\hat{b}_k^{\dag})/\sqrt{2}$.
Then, the Wigner function corresponding to ${\rho}_{env}$ of Eq.(\ref{rhoenv}) reads
 \begin{eqnarray}
W_{env}(\mbox{\boldmath$v$})=\frac{1}{\pi^2}
\exp\left[-\mbox{\boldmath$v$}{\cal A}_s \mbox{\boldmath$v$}^T\right]\,,
\end{eqnarray}
where ${\cal A}_s$ is given by Eq.(\ref{Ar}) with the replacement $r\to s$.

To go further on, let us define the vectors in $\mathbb{R}^{8}$ 
\begin{equation}
\mbox{\boldmath$\gamma$}=(\mbox{\boldmath$u$},\mbox{\boldmath$v$})\,,\quad
\mbox{\boldmath$\theta$}=(\mbox{\boldmath$u$},\mbox{\boldmath$0$})\,,\quad
\mbox{\boldmath$\tau$}=(\mbox{\boldmath$0$},\mbox{\boldmath$v$})\,,\quad
\mbox{\boldmath$\kappa$}=(\mbox{\boldmath$\mu$},\mbox{\boldmath$0$})\,.
\label{gathtaka}
\end{equation}
We can write the total (input plus environment) Wigner function,
corresponding to  ${\rho}_{in}\otimes {\rho}_{env}$ as
\begin{eqnarray}\label{Wtot}
&&W_{in}\left(\mbox{\boldmath$u$};\mbox{\boldmath$\mu$}\right)W_m\left(\mbox{\boldmath$v$}\right)\\
&=&\frac{1}{\pi^4}
\exp\left[
-\mbox{\boldmath$\gamma$}{\cal A}\mbox{\boldmath$\gamma$}^T+2\mbox{\boldmath$\kappa$}{\cal A}\mbox{\boldmath$\gamma$}^T-\mbox{\boldmath$\kappa$}{\cal A}\mbox{\boldmath$\kappa$}^T\right]\,,
\nonumber
\end{eqnarray}
where ${\cal A}$ is the $8 \times 8$ diagonal block matrix
\begin{equation}
{\cal A}=
\left(
\begin{array}{cc}
{\cal A}_r  &  0  \\
0  &  {\cal A}_s   
\end{array}
\right)\,.
\end{equation}

As consequence of Eqs.(\ref{unouno}), (\ref{duedue}) the signal-noise coupling  
 corresponds to the change of variables 
\begin{equation}\label{bs}
\mbox{\boldmath$\gamma$}^T\longrightarrow {\cal B}\mbox{\boldmath$\gamma$}^T\,
\end{equation}
produced by the unitary beam splitter matrix
\begin{equation}
\label{O}
{\cal B}=
\left(
\begin{array}{ccc}
\sqrt{\eta}\,\, {\cal I} &    \sqrt{1-\eta} \,\,  {\cal I}  \\
 -\sqrt{1-\eta} \,\,  {\cal I}  &  \sqrt{\eta} \,\,  {\cal I} 
\end{array}
\right)\,,
\end{equation}
with $ {\cal I}$ the $4\times 4$ identity matrix.
By using Eq.(\ref{bs}) into (\ref{Wtot}) we obtain the total Wigner function after the interaction between input and environment, thus corresponding to the state $\rho$ of Eq.(\ref{rhomap})
\begin{align}
W(\mbox{\boldmath$\gamma$}; \mbox{\boldmath$\kappa$})=\frac{1}{\pi^4}
\exp&\left[
-\mbox{\boldmath$\gamma$}\,  {\cal B}^T{\cal A}{\cal B}  \,\mbox{\boldmath$\gamma$}^T\right.\nonumber\\
&\left.+2\mbox{\boldmath$\kappa$}\,  {\cal A}{\cal B} \,\mbox{\boldmath$\gamma$}^T-\mbox{\boldmath$\kappa$}\,{\cal A}\,\mbox{\boldmath$\kappa$}^T
\right].
\label{Wtotafter}
\end{align}


\section{Private Classical Information Capacity}

Integrating Eq.(\ref{Wtotafter}) over the variable $\mbox{\boldmath$v$}$ (resp. $\mbox{\boldmath$u$}$), we get the output (resp. eavesdropper) Wigner function, $W_{out}(\mbox{\boldmath$\theta$}; \mbox{\boldmath$\kappa$})$ (resp. $W_{eve}(\mbox{\boldmath$\tau$}; \mbox{\boldmath$\kappa$})$).
It means to have the following correspondences 
\begin{eqnarray}
{\rho}_{out}&\leftrightarrow&
W_{out}(\mbox{\boldmath$\theta$};\mbox{\boldmath$\kappa$})=\int d\mbox{\boldmath$v$} \;W(\mbox{\boldmath$\gamma$};\mbox{\boldmath$\kappa$}),\label{rhoout}\\
{\rho}_{eve}&\leftrightarrow&
W_{eve}(\mbox{\boldmath$\tau$};\mbox{\boldmath$\kappa$})=\int d\mbox{\boldmath$u$} \;W(\mbox{\boldmath$\gamma$};\mbox{\boldmath$\kappa$}).\label{rhoeve}
\end{eqnarray}
Averaging over the input distribution (\ref{Pimu}), we also have
\begin{eqnarray}
{\overline\rho}_{out}&\leftrightarrow&
{\overline W}_{out}(\mbox{\boldmath$\theta$})=\int d\mbox{\boldmath$\mu$} \;P(\mbox{\boldmath$\mu$}){W}_{out}(\mbox{\boldmath$\theta$};\mbox{\boldmath$\kappa$}),
\label{orhoout}\\
{\overline\rho}_{eve}&\leftrightarrow&
{\overline W}_{eve}(\mbox{\boldmath$\tau$})=\int d\mbox{\boldmath$\mu$} \;
P(\mbox{\boldmath$\mu$}){W}_{eve}(\mbox{\boldmath$\tau$};\mbox{\boldmath$\kappa$}).
\label{orhoeve}
\end{eqnarray}

From Eqs.(\ref{rhoout}), (\ref{rhoeve}), (\ref{orhoout}), (\ref{orhoeve}) taking into account Eq.(\ref{gathtaka}), we get the following Gaussian functions
\begin{eqnarray}
W_{out}(\mbox{\boldmath$u$};\mbox{\boldmath$\mu$})&=&\frac{1}{(2\pi)^2\sqrt{\det V_{out}}}
\exp \left[-\frac{1}{2} \mbox{\boldmath$u$} V_{out}^{\,-1}\mbox{\boldmath$u$}^T\right.\\
 & & \left.+\sqrt{\eta}\mbox{\boldmath$\mu$}V_{out}^{\,-1}\mbox{\boldmath$u$}^T
-\frac{1}{2}\eta \mbox{\boldmath$\mu$}V_{out}^{\,-1}\mbox{\boldmath$\mu$}^T\right],\nonumber\\
W_{eve}(\mbox{\boldmath$v$};\mbox{\boldmath$\mu$})&=&\frac{1}{(2\pi)^2\sqrt{\det V_{eve}}} \exp\left[-\frac{1}{2}
\mbox{\boldmath$v$} V_{eve}^{\,-1}\mbox{\boldmath$v$}^T\right.\\
 & & \left.+\sqrt{1-\eta}\mbox{\boldmath$\mu$}V_{eve}^{\,-1}\mbox{\boldmath$v$}^T
-\frac{1}{2}(1-\eta) \mbox{\boldmath$\mu$}V_{eve}^{\,-1}\mbox{\boldmath$\mu$}^T\right],\nonumber\\
{\overline W}_{out}(\mbox{\boldmath$u$})&=&\frac{1}{(2\pi)^2\sqrt{\det{\overline V}_{out}}}\exp{\left[-\frac{1}{2}
\mbox{\boldmath$u$} {\overline V}_{out}^{\,-1}\mbox{\boldmath$u$}^T
\right]},\\
{\overline W}_{eve}(\mbox{\boldmath$v$})&=&\frac{1}{(2\pi)^2\sqrt{\det{\overline V}_{eve}}}\exp{\left[-\frac{1}{2}
\mbox{\boldmath$v$} {\overline V}_{eve}^{\,-1}\mbox{\boldmath$v$}^T
\right]},
\end{eqnarray}
whose  covariance matrices result
\begin{eqnarray}
V_{out}\,&=&\frac{1}{2}\left[\eta {\mathcal A}_{r}^{-1}+\left(1-\eta\right){\mathcal A}_{s}^{-1}\right],\label{Vout}\\
V_{eve}\,&=&\frac{1}{2}\left[\left(1-\eta\right) {\mathcal A}_{r}^{-1}+\eta {\mathcal A}_{s}^{-1}\right],\label{Veve}\\
{\overline V}_{out}&=&V_{out}+\frac{1}{2}\,\eta\, N {\cal I},\label{oVout}\\
{\overline V}_{eve}&=&V_{eve}+\frac{1}{2}\left(1-\eta\right) \,N {\cal I}.\label{oVeve}
\end{eqnarray}

Since any excess information the receiver has relative to eavesdropper can in principle be exploited by receiver and sender to distill a shared secret key \cite{CK78}, it makes sense to consider the
difference between output and eavesdropper Holevo's information 
as guaranteed privacy of the channel \cite{Dev}.
Hence, we introduce the private information normalized to the number of channel uses 
\begin{eqnarray}
\label{Ip}
I_p(2)=\frac{1}{2}\left(\chi_{out}-\chi_{eve}\right),
\end{eqnarray}
with
\begin{eqnarray}
\chi_{out}&=& S\left(\overline{\rho}_{out}\right)-\int d\mbox{\boldmath$\mu$} P(\mbox{\boldmath$\mu$})
S\left({\rho}_{out}\right),\label{chiout}\\
\chi_{eve}&=& S\left(\overline{\rho}_{eve}\right)-\int d\mbox{\boldmath$\mu$} P(\mbox{\boldmath$\mu$})
S\left({\rho}_{eve}\right),\label{chieve}
\end{eqnarray}
being $S$ the von Neumann entropy.
The supremum of private information (\ref{Ip}) represents the 2-shot private classical information capacity \cite{Dev}. Due to the initial conjecture about optimality of Gaussian encoding and the generality of the state (\ref{psi}), it amounts to the maximum of $I_p$ over parameter $r$.

The symplectic eigenvalues of covariance matrices 
(\ref{Vout}), (\ref{Veve}), (\ref{oVout}), (\ref{oVeve})
\begin{eqnarray}
\lambda_{out,j}&=&\frac{1}{2}\left[1-2\,\eta\,\left(1-\eta\right)+2\,\eta\,\left(1-\eta\right)\,\cosh 2(r-s)\right]^{1/2},\nonumber\\
 & & \\
\lambda_{eve,j}&=&\frac{1}{2}\left[1-2\,\eta\,\left(1-\eta\right)+2\,\eta\,\left(1-\eta\right)\,\cosh 2(r-s)\right]^{1/2},\nonumber\\
 & & \\
{\overline\lambda}_{out,j}&=&\frac{1}{2}\left\{1+\eta^2\left(N^2+2\,N\cosh 2r\right)\right.\nonumber\\
& & \left.+2\,\eta\,(1-\eta)\left[\cosh 2(r-s)+N\cosh 2s-1\right]\right\}^{1/2},\nonumber\\
& & \\
{\overline\lambda}_{eve,j}&=&\frac{1}{2}\left\{1+(1-\eta)^2\left(N^2+2\,N\cosh 2r\right)\right.\nonumber\\
& & \left.+2\,\eta\,(1-\eta)\left[\cosh 2(r-s)+N\cosh 2s-1\right]\right\}^{1/2},\nonumber\\
& & ,
\end{eqnarray}
allow to calculate the entropies in Eqs.(\ref{chiout}) and (\ref{chieve}) as \cite{Hol99}
\begin{eqnarray}
S({\rho}_{out})&=&\sum_{j=1}^{2}g\left(|\lambda_{out,j}|-\frac{1}{2}\right),
\label{Srhoout}\\
S({\rho}_{eve})&=&\sum_{j=1}^{2}g\left(|\lambda_{eve,j}|-\frac{1}{2}\right),
\label{Srhoeve}\\
S(\overline{\rho}_{out})&=&\sum_{j=1}^{2}g\left(|{\overline\lambda}_{out,j}|-\frac{1}{2}\right),\\
S(\overline{\rho}_{eve})&=&\sum_{j=1}^{2}g\left(|{\overline\lambda}_{eve,j}|-\frac{1}{2}\right),
\end{eqnarray}
where $g(x)=(x+1)\log_2 (x+1)-x\log_2 x$.

It results that Eqs.(\ref{Srhoout}) and (\ref{Srhoeve}) do not depend on $\mbox{\boldmath$\mu$}$, hence
we straightforward get $\chi_{out}$ and $\chi_{eve}$ of Eqs.(\ref{chiout}) and (\ref{chieve}) since  both integrals amount to 1.


\section{Results and Conclusions}

We are now in the position to analyze the behavior of the quantity $I_p$ of Eq.(\ref{Ip}).
Since we want to bound the effective average photon number per channel use, in practice we fix $N_{eff}$ as the effective input photon number and we consider $N$ varying as function of $r$ ($N=N_{eff}-\sinh^2 r$), limiting the range of $r$ to those values for which $N\ge 0$.

\begin{figure}
\centering
\includegraphics[width=2.7in]{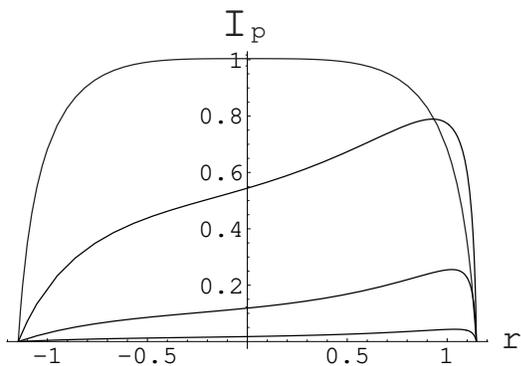}
\caption{Private information $I_p$ versus the entanglement parameter $r$.
Curves from top to bottom are for $s=0$, $1$, $2$, $3$. 
The values of other parameters are $\eta=0.8$ and $N_{eff}=2$.}
\label{fig1}
\end{figure}

\begin{figure}
\centering
\includegraphics[width=2.7in]{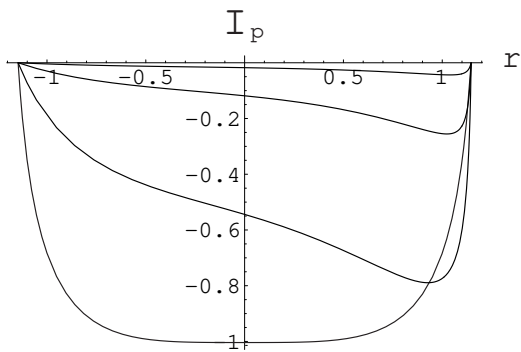}
\caption{Private information $I_p$  versus the entanglement parameter $r$.
Curves from bottom to top are for $s=0$, $1$, $2$, $3$. 
The values of other parameters are $\eta=0.2$ and $N_{eff}=2$.}
\label{fig2}
\end{figure}

Another parameter to take into account is the beam splitter transmittivity $\eta$. We distinguish three limit cases.  
For $\eta=1$, the channel is not a lossy channel, thus the whole information arrives to the receiver; i.e. $I_p\ge 0$ and it is maximum with respect to $\eta$. For $\eta=1/2$ the receiver and the eavesdropper have the same information; i.e. $I_p=0$. For $\eta=0$ the whole information is lost and captured by the eavesropper; i.e. $I_p\le 0$ and it is minimum with respect to $\eta$. 
Here we present two intermediate situations: $\eta=0.8$ (Fig.\ref{fig1}) and $\eta=0.2$ (Fig.\ref{fig2}). 
In Fig.\ref{fig1} the private information $I_p$ is shown {\em vs.} the entanglement parameter $r$ for different values of the degree of memory $s$, for an input photon number $N_{eff}=2$. 
For $s=0$ the behavior of $I_p$ is symmetric with respect to the value $r=0$, at which it attains the maximum. That is, entangled inputs are no way useful in the memoryless case. 
As soon as the degree of memory increases ($s>0$) the symmetry is broken. For a certain range of $r$'s values the function $I_p$ goes above its  value corresponding to product input state ($r=0$). This clearly shows an improvement of the security of transmission  through a memory channel, due to entangled inputs instead of product states. However, the effectiveness of entangled inputs in presence of memory never overcomes that of unentangled inputs in absence of memory (the maxima of the curves for $s> 0$ are below the maximum of the curve for $s=0$). This is in contrast with what happen for the simple classical capacity \cite{Rug05,Cerf04}.

It is also intuitive that by increasing the memory strength, the private information becomes lower and lower until flattening to zero. In fact, correlations among the (classical) symbols help the eavesdropper in predicting them, while completely uncorrelated symbols would avoid that.

In Fig.\ref{fig2} the privacy information {\em vs.} the entanglement parameter $r$ is negative, as it is expected for $\eta\le 1/2$.\footnote{$I_p$ shows a mirror symmetry with respect to abscissa when $\eta\rightarrow (1-\eta)$ and a mirror symmetry with respect to ordinate when $s\rightarrow -s$.} Hence, in such cases the channel does  not  represent a secure mean for information transmission even in presence of memory. 
The actual security threshold $\eta=1/2$ recalls those found for secure continuous variable cryptographic key distribution \cite{Gro05}. 
In particular, our model corresponds to lossy channel coherent attack where also the receiver is allowed to perform general collective measurements. Thus, 
due to the symmetry between receiver's and eavesdropper's operations, the security threshold coincides with that of individual attacks $\eta=1/2$ \cite{Gro05}.  

It is straightforward to extend the studied model to many uses of the channel, 
by employing the same correlation strength among all environment modes (like in Ref.\cite{Rug05}),
that is by employing an operator $\hat{S}_b$ of the form
$\exp\left[\frac{1}{2}s\sum_{k\neq k'}\left(\hat{b}_k^{\dag}\hat{b}_{k'}^{\dag}-\hat{b}_k\hat{b}_{k'}\right)\right]$. In such a case
we can observe that the quantities $\chi_{out}$ and  $\chi_{eve}$ are linear in the number of uses (number of modes). Thus, $I_p(n)=I_p(2)$ and the presented results still hold. A much more demanding task would be the study  the private classical information when non symmetric noise correlations are involved in many channel uses, i.e. when the memory has a finite range over channel uses. 
However, we forecast the same conclusions about the effectiveness of entangled inputs.
   
In conclusion, we have studied the private classical information capacity for a lossy bosonic channel including memory effects. We have shown the possibility to enhance it by means of entangled inputs with respect to product states. However, the effectiveness of entangled inputs in presence of memory never overcomes that of unentangled inputs in absence of memory. This is in contrast with what happen for simple classical capacity.


\end{document}